\documentclass[prl,notitlepage,superscriptaddress,showpacs,amsmath,amssymb,twocolumn]{revtex4-1}
\usepackage{hyperref,graphicx}
\usepackage{xcolor}

\usepackage{siunitx}

\begin{document}

 \title{Dynamical separation of bulk and edge transport in HgTe-based 2D topological insulators}

\author{Matthieu C. Dartiailh}
\affiliation{Laboratoire de Physique de l'\'Ecole Normale Sup\'erieure, ENS, PSL Research University, CNRS, Sorbonne Universit\'e, Universit\'e Paris Diderot, Sorbonne Paris Cit\'e, 24 rue Lhomond, 75231 Paris Cedex 05, France}
\author{Simon Hartinger}
\affiliation{Physikalisches Institut (EP3), Am Hubland, Universit\"at W\"urzburg, D-97074 W\"urzburg, Germany}
\affiliation{Institute for Topological Insulators, Am Hubland, Universit\"at W\"urzburg, D-97074 W\"urzburg, Germany}
\author{Alexandre Gourmelon}
\affiliation{Laboratoire de Physique de l'\'Ecole Normale Sup\'erieure, ENS, PSL Research University, CNRS, Sorbonne Universit\'e, Universit\'e Paris Diderot, Sorbonne Paris Cit\'e, 24 rue Lhomond, 75231 Paris Cedex 05, France}
\author{Kalle Bendias}
\affiliation{Physikalisches Institut (EP3), Am Hubland, Universit\"at W\"urzburg, D-97074 W\"urzburg, Germany}
\affiliation{Institute for Topological Insulators, Am Hubland, Universit\"at W\"urzburg, D-97074 W\"urzburg, Germany}
\author{Hugo Bartolomei}
\author{Hiroshi Kamata}
\author{Jean-Marc Berroir}
\author{Gwendal F\`eve}
\author{Bernard Pla\c cais}
\affiliation{Laboratoire de Physique de l'\'Ecole Normale Sup\'erieure, ENS, PSL Research University, CNRS, Sorbonne Universit\'e, Universit\'e Paris Diderot, Sorbonne Paris Cit\'e, 24 rue Lhomond, 75231 Paris Cedex 05, France}
\author{Lukas Lunczer}
\author{Raimund Schlereth}
\author{Hartmut Buhmann}
\author{Laurens W. Molenkamp}
\affiliation{Physikalisches Institut (EP3), Am Hubland, Universit\"at W\"urzburg, D-97074 W\"urzburg, Germany}
\affiliation{Institute for Topological Insulators, Am Hubland, Universit\"at W\"urzburg, D-97074 W\"urzburg, Germany}
\author{Erwann Bocquillon}
\affiliation{Laboratoire de Physique de l'\'Ecole Normale Sup\'erieure, ENS, PSL Research University, CNRS, Sorbonne Universit\'e, Universit\'e Paris Diderot, Sorbonne Paris Cit\'e, 24 rue Lhomond, 75231 Paris Cedex 05, France}
\email{erwann.bocquillon@lpa.ens.fr}

\date{\today}

\begin{abstract}
Topological effects in edge states are clearly visible on short lengths only, thus largely impeding their studies. On larger distances, one may be able to dynamically enhance topological signatures by exploiting the high mobility of edge states with respect to bulk carriers. Our work on microwave spectroscopy highlights the responses of the edges which host very mobile carriers, while bulk carriers are drastically slowed down in the gap. Though the edges are denser than expected, we establish that charge relaxation occurs on short timescales, and suggests that edge states can be addressed selectively on timescales over which bulk carriers are frozen.
\end{abstract}


\maketitle

{\bf
}

Promising platforms have emerged to investigate the physics of two-dimensional topological insulators, which exhibit the quantum spin Hall (QSH) effect. The most prominent examples are HgTe quantum wells, in which many transport signatures of the QSH effect have been observed \cite{konig2007,roth2009, brune2012,calvo2017} as well as a fractional Josephson effect, a signature of topological superconductivity \cite{bocquillon2016,deacon2017}, in HgTe-based Josephson junctions. Another promising material system, InAs/GaSb double quantum well, also shows the quantized conductance that accompanies topological edge state transport \cite{knez2011,du2015}. Several other layered materials are also currently under development, such as bismuthene \cite{reis2017} or WTe$_2$ \cite{wu2018}. However, the progress in transport studies of the QSH effect has been severely hampered as signatures of topology are at most visible on a few microns despite the expected topological protection \cite{konig2007,du2015,bendias2018}. 
A significant reason is the presence of bulk bands, which naturally form puddles near the gap. Though these states poorly conduct, they may introduce scattering of the topological states \cite{lunczer2019}. The associated density of states (DOS) is large, especially in HgTe QWs with indirect band gap. 

Topological signatures may be enhanced in dynamical studies by exploiting the difference in transport or scattering timescales between topological and bulk carriers. Recent efforts have for example demonstrated that dynamical experiments in the THz range \cite{luo2019} or using time-resolved ARPES \cite{soifer2019} allow the investigation and manipulation of sensitive topological states despite bulk background carriers. Transposing this idea to a microwave transport regime, we propose microwave drives to dynamically isolate quasi-ballistic topological edge modes while bulk carriers have an evanescent contribution since their diffusivity collapses near the gap of the material.

Here, we report on a systematic study of microwave capacitance spectroscopy, in the frequency range \SI{10}{\kilo\hertz}--\SI{10}{\giga\hertz}, to investigate narrow HgTe quantum wells (QW). Microwave ac signals enable simultaneous measurements of the capacitive response, sensitive to the density of electronic states, and of the resistive response, probing their ability to conduct, as exemplified recently in three-dimensional topological insulators \cite{xu2015,kozlov2016,inhofer2017,inhofer2018}. Microwave capacitance spectroscopy thus naturally evidences the resulting $RC$ charge relaxation times of each transport channel. We combine it with geometric scalings and gate control of the electron density.
In this article, we show that the conduction and valence bands of the QW are ruled by a single-mode dynamics, corresponding to bulk carriers in good agreement with ${\bf k\cdot p}$ band structure predictions. In contrast, the microwave spectra confirms a striking two-mode dynamics when the Fermi level approaches the gap. The dynamical transport properties can be ascribed to bulk and edge carriers, which is confirmed by geometric scalings \cite{sm_prl}. As expected, the edges host very mobile carriers but they appear also much denser than predicted for bare helical edge states. It suggests that the structure of topological edges is complex for example due to electrostatics and/or disorder. In contrast, bulk states have high resistance and contribute an enormous density of states (DOS) near the gap, resulting in a slow dynamics. Consequently, the charge relaxation frequencies associated to bulk and edge carriers differ by more than one order of magnitude, thus paving the way for more robust experiments in which topological edges can be dynamically and selectively addressed, while bulk carriers remain frozen.

\begin{figure}[ht]
\centerline{\includegraphics[width=0.5\textwidth]{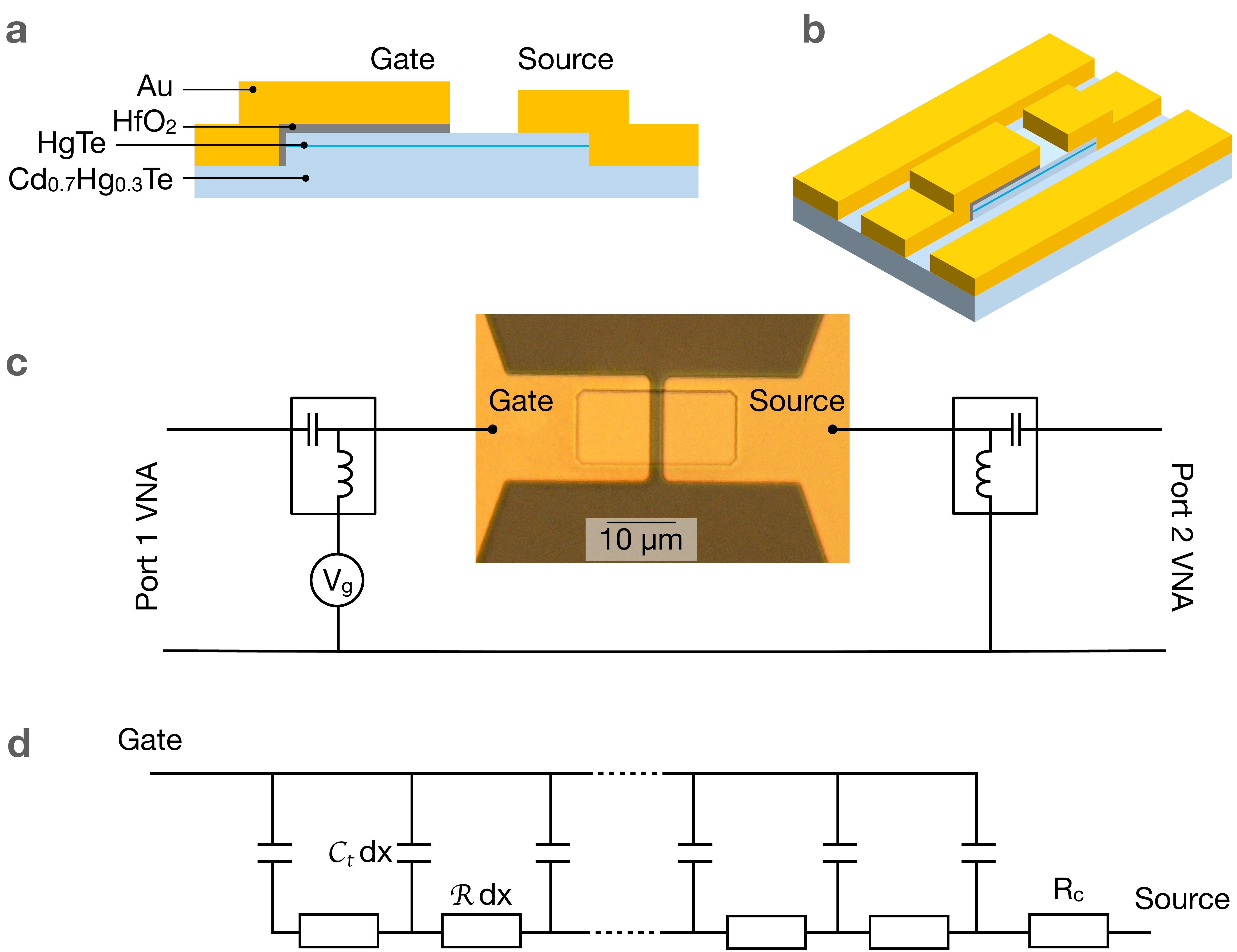}}
\caption{{\bf Experimental techniques:} a) Schematic side view of the device, showing the different layers of the CdHgTe/HgTe heterostructure, as well as the gate and contact. b) Artist view of the capacitor device, embedded in a coplanar waveguide c) High-frequency measurement setup, based on a vector network analyzer (VNA), the ports of which are connected to both source and gate. Two bias-tees are used to apply the dc gate voltage $V_{\rm g}$ that controls the electron density. For low frequency measurements, a lock-in replaces the VNA. d) Model of the capacitor as a distributed RC line, with a contact resistance $R_{\rm c}$, and a linear resistance $\mathcal{R}$ and linear capacitance $\mathcal{C}_{\rm t}$.
}\label{figure:sample}
\end{figure}

\begin{figure}[ht]
\centerline{\includegraphics[width=8.5cm]{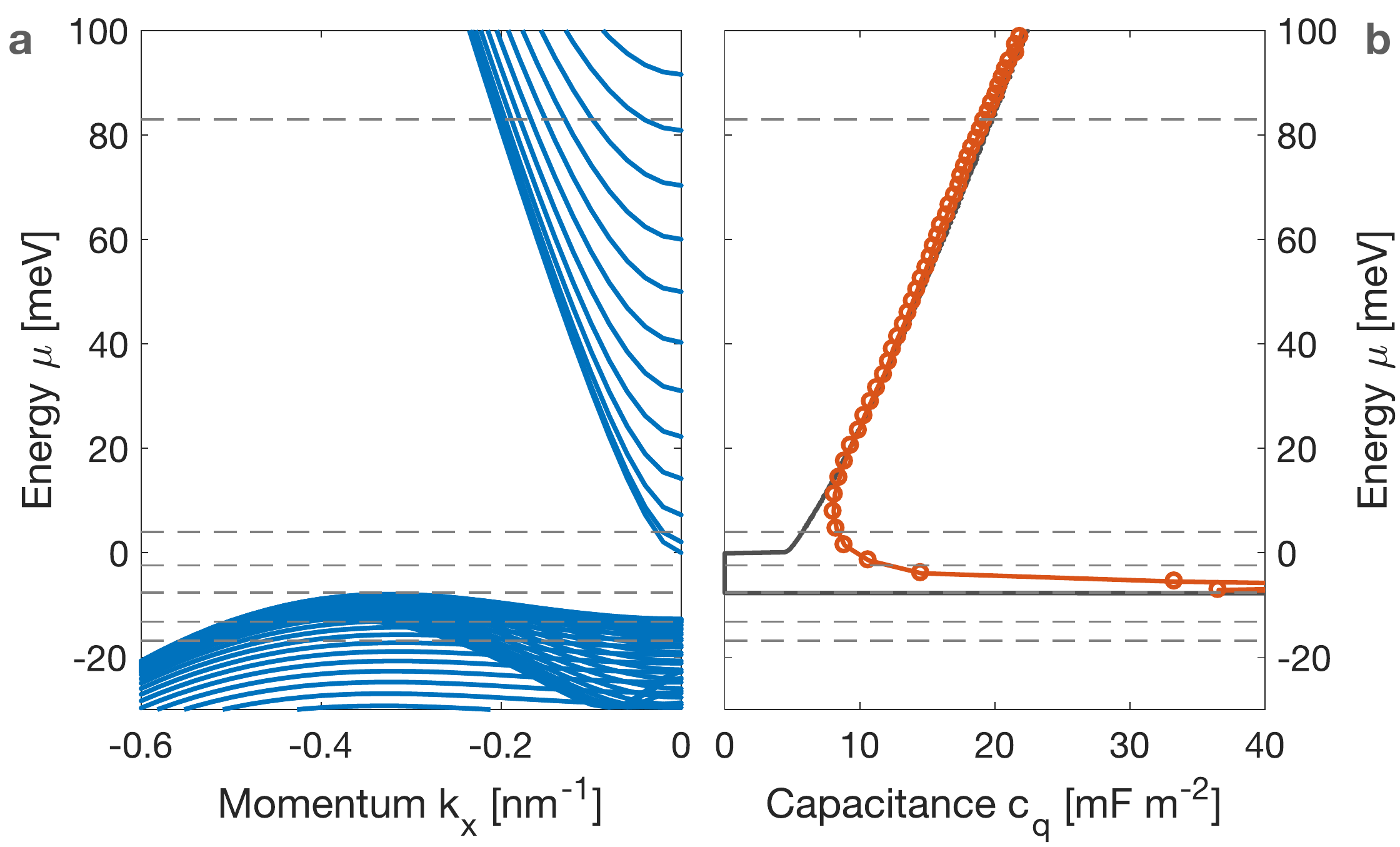}}
\caption{{\bf Quantum capacitance of HgTe quantum wells:} a) Energy bands as function of momentum $k_x$, as calculated for different values of transverse momentum $k_y$. b) Quantum capacitance $c_{\rm q}$ (per unit area) as function of energy $\mu$ (for a size of $L=\SI{10}{\micro\meter}$). The grey line corresponds to the computed value of $c_{\rm q}$ given the ${\bf k\cdot p}$ band structure, the red dots and line to experimental data.
}\label{figure:BS}
\end{figure}

\paragraph{Samples --}
Our samples are based on HgTe/HgCdTe QWs grown by molecular beam epitaxy on CdTe substrates (see Fig.\ref{figure:sample}a). The thickness $t$ of the quantum wells varies between 5 and \SI{11}{\nano\meter}, while the protective HgCdTe capping layers has a typical thickness of \SI{15}{\nano\meter}. In these quantum wells, the band structure consists of light electrons in the conduction band, and heavier holes in the valence band. A topological phase transition occurs for the critical thickness $t_c\simeq \SI{6.6}{\nano\meter}$. For $t>t_c$ an inverted band ordering is responsible for the appearance of topologically protected quantum spin Hall edge states in the gap of the quantum well \cite{bernevig2006}. For $t<t_c$, the band ordering is normal and no edge states occur, the sample is topologically trivial. The quantum wells are first characterized using standard Hall-bar measurements, which yield a typical mobility around \SI{2e5}{\square\centi\meter\per\volt\per\second} (measured at a density \SI{3e11}{\per\square\centi\meter} in the conduction band). A series of QWs have been investigated and have given similar results. This letter focuses on several devices realized on one layer, with a thickness of \SI{8}{\nano\meter} and a predicted gap around \SI{8}{\milli\electronvolt} as shown in Fig.\ref{figure:BS}a. Additional data obtained on other topological layers, and control data on a trivial (non-inverted) sample are reported in the Supplementary Materials \cite{sm_prl}. The devices consist of square capacitors of side $L=3, 5, 10, \SI{20}{\micro\meter}$, embedded in gold coplanar waveguides ( Fig.\ref{figure:sample}a and \ref{figure:sample}b). For these devices, the capacitor mesa is defined via a wet etching technique \cite{bendias2018} that preserves the high crystalline quality and the high mobility of the epilayer. The contacts and gate  are patterned via optical lithography. The gold gate electrode is evaporated on top of a ~\SI{7}{\nano\meter}-thick HfO$_2$ insulating layer, grown by low-temperature atomic layer deposition (ALD)\cite{bendias2018}. An ohmic contact is deposited by Au evaporation.

\paragraph{Admittance measurements --}

The capacitors are measured in a cryogenic RF probe station at a temperature $T=\SI{10}{\kelvin}$ using a vector network analyzer (VNA) over three decades of frequency $\SI{10}{\mega\hertz}<f<\SI{10}{\giga\hertz}$. Standard in-situ calibration techniques enable to de-embed the response of the circuitry and stray capacitances from the admittance $Y(\omega)$ of interest, with $\omega=2\pi f$. Earlier works \cite{pallecchi2011,inhofer2017,inhofer2018} have demonstrated that the capacitors can be (assuming translation invariance along the transverse direction) described by a distributed RC line of length $L$, with line capacitance $\mathcal{C}_{\rm t}$ and resistance $\mathcal{R}$, such as the one depicted in Fig.\ref{figure:sample}d. At the lowest order in frequency, the device is equivalent to a capacitor $C_{\rm t}=\mathcal{C}_{\rm t}L$, with the admittance given by $Y(\omega)=i\omega C_{\rm t}$. The total capacitance $C_{\rm t}$ is the series addition of the standard geometric capacitance $C_{\rm g}$ and the quantum capacitance $C_{\rm q}$, which accounts for the increase of the Fermi energy and reads $C_{\rm q}=e^2\rho$, where $\rho$ is the DOS (for weakly interacting systems in the limit of zero temperature). The geometric capacitance $C_{\rm g}$ is extracted at high densities in the valence band for which $C_{\rm t}\simeq C_{\rm g}$ (see \cite{pallecchi2011,inhofer2017,sm_prl}). Since the QW is never strongly depleted even in the gap, the geometry remains that of a plate capacitor. We thus calculate $C_{\rm q}$ assuming a constant $C_{\rm g}$. Importantly, we compute the chemical potential \cite{berglund1966} $\mu=\int^{V_{\rm g}} \left(1-\frac{C_{\rm t}(V)}{C_{\rm g}}\right){\rm d}V$ to allow for a direct comparison with predictions obtained from ${\bf k\cdot p}$ band structure calculations \cite{novik2005}. The origin of the chemical potential is fixed to the maximum of the sample sheet resistance\cite{sm_prl}.

\paragraph{Quantum capacitance --}
Fig.\ref{figure:BS} shows experimental data of the (areal) quantum capacitance $c_{\rm q}(\mu)~=~C_{\rm q}(\mu)/L^2$ and ${\bf k\cdot p}$ predictions. The capacitance is low in the conduction band ($c_{\rm q}<\SI{20}{\milli\farad\per\square\meter}$ for $\mu>0$), and increases quasi-linearly with $\mu$, as expected for two-dimensional bands with quasi-linear band dispersion. The agreement with ${\bf k\cdot p}$ calculations is excellent in this regime, provided the value of $C_{\rm g}$ is fine tuned (within $\pm 2\%$). In this regime, the quantum capacitance is given by \cite{inhofer2017} $c_{\rm q}=\frac{e^2}{\pi(\hbar v_{\rm F})^2}\mu$, with the Fermi velocity $v_{\rm F}\simeq\SI{1e6}{\meter\per\second}$. In the valence band ($\mu\lesssim 0$), we observe that $c_{\rm q}$ is very high ($c_{\rm q}>\SI{40}{\milli\farad\per\square\meter}$) as the valence band has a high mass. $c_{\rm q}$ is however difficult to extract with high precision in this range, due to the finite value of the geometric capacitance (here $c_{\rm g}\simeq\SI{3.91}{\milli\farad\per\square\meter}$).
In between these two regimes, a minimum is observed in the data $c_{\rm q}\simeq\SI{10}{\milli\farad\per\square\meter}$. It indicates the gap of the QW, where QSH edge states should be observed. A modeling as merely one-dimensional edge states yields a quantum capacitance per unit length of $\frac{4e^2}{hv_{\rm F}}\simeq \SI{0.2}{\nano\farad\per\meter}$ for the topological edge states \cite{chaste2008,entin2017}. On a $L=\SI{10}{\micro\meter}$ device, it corresponds to a very low areal contribution $c_{\rm q}<\SI{0.1}{\milli\farad\per\square\meter}$, much smaller than the observed minimum. This residual contribution in the gap is thus unlikely to originate solely from the edge states, but more likely from disorder in the bulk bands. The smearing is much larger than the temperature ($kT\lesssim \SI{1}{\milli\electronvolt}$) and any broadening induced by the probe signal, and may be attributed to inhomogeneous bulk bands, charge puddles due to disorder, thickness fluctuations \cite{tkachov2011} or electrostatics. As a consequence of the extremely large DOS of the valence band, the minimum of $c_{\rm q}$ is surprisingly up-shifted in energy with respect to the theoretically predicted gap (here typically around $\mu=\SI{10}{\milli\electronvolt}$).

\paragraph{Edge and bulk dynamics --}
For higher frequencies, dissipative and propagative effects set in. As depicted in Fig.\ref{figure:sample}d, dissipation results from the access resistance $R_{\rm c}$ and the finite sheet resistance $R=\mathcal{R}L$ of the HgTe.  For frequencies exceeding the charge relaxation frequency $1/2\pi RC_t$ of the device, typically \SI{1}{\giga\hertz} for $L=\SI{10}{\micro\meter}$, the capacitors then host evanescent waves \cite{pallecchi2011, inhofer2017, graef2018b}, driven by the resistance of the HgTe film. Assuming translation invariance along the transverse axis, the admittance of the capacitor reads \cite{sm_prl}:
\begin{equation}
\label{eq:admittance}
Y(\omega)=\frac{1}{R_{\rm c}+ \frac{R}{ik\tanh\left(ik\right)}}
\end{equation}
where $R_{\rm c}$ describes the lumped contact resistance, while the $\tanh$ term describes the evanescent waves in the capacitor, with $k=\sqrt{iRC_{\rm t}\omega}$.

\label{section:hf}

\begin{figure*}[ht]
\centerline{\includegraphics[width=\textwidth]{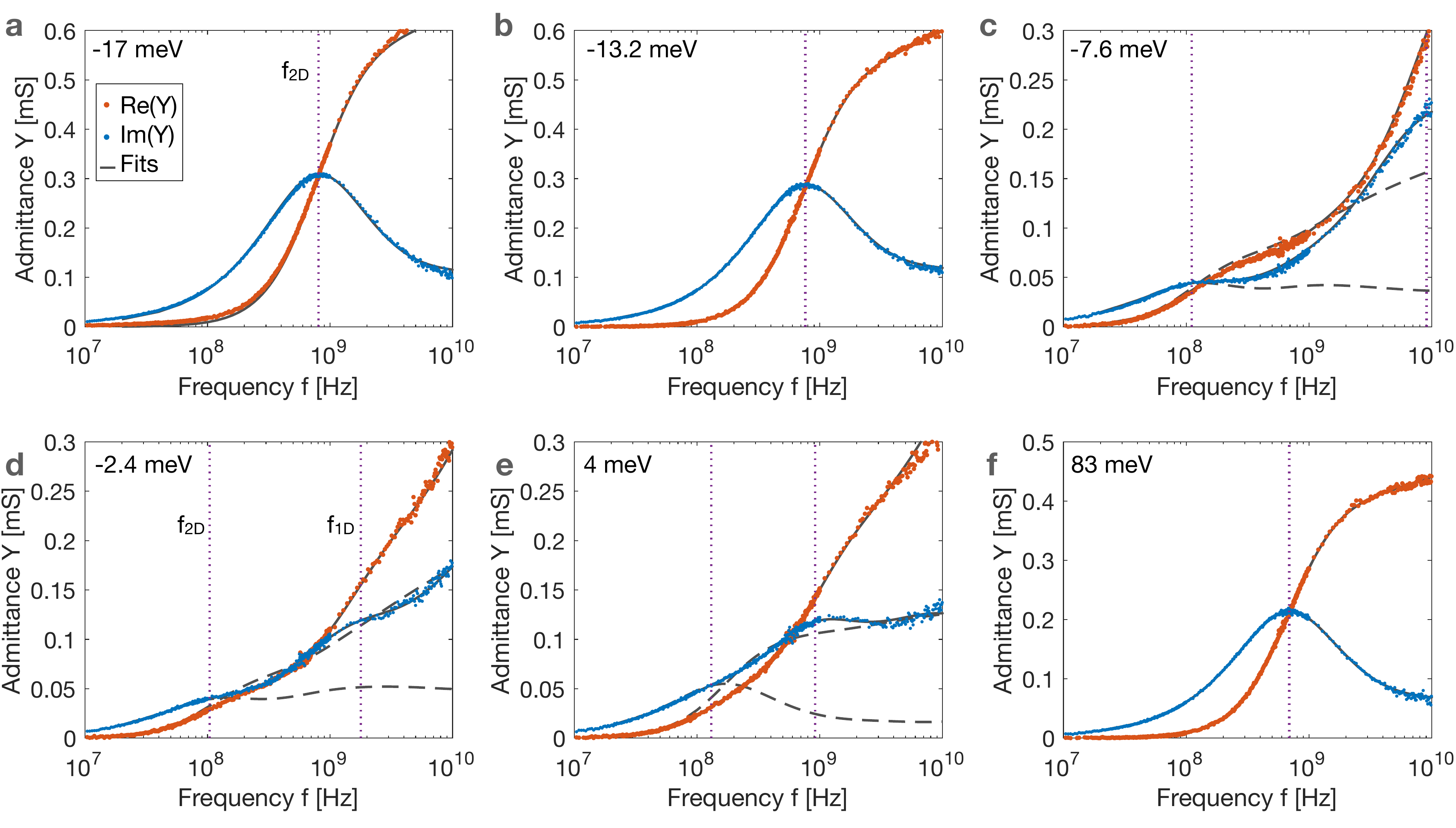}}
\caption{{\bf High-frequency admittance:} Admittance $Y$ as function of frequency $f$ for different energies $\mu$ (indicated in each panel, see also Fig.\ref{figure:BS}) for $L=\SI{5}{\micro\meter}$. Panels a, b correspond to the valence band, f to the conduction band, and show a good agreement with a single-mode model (plain grey line). The other panels (c, d, e) correspond to the transition from valence to conduction bands. They are only well fitted by the two-mode model (plain grey line), while single-mode models fail (dashed grey line). The vertical dotted lines indicate the values of the charge relaxation frequencies $f_{\rm 2D}$ and $f_{\rm 1D}$.
}\label{figure:highfreq}
\end{figure*}

\begin{figure}[ht]
\centerline{\includegraphics[width=0.5\textwidth]{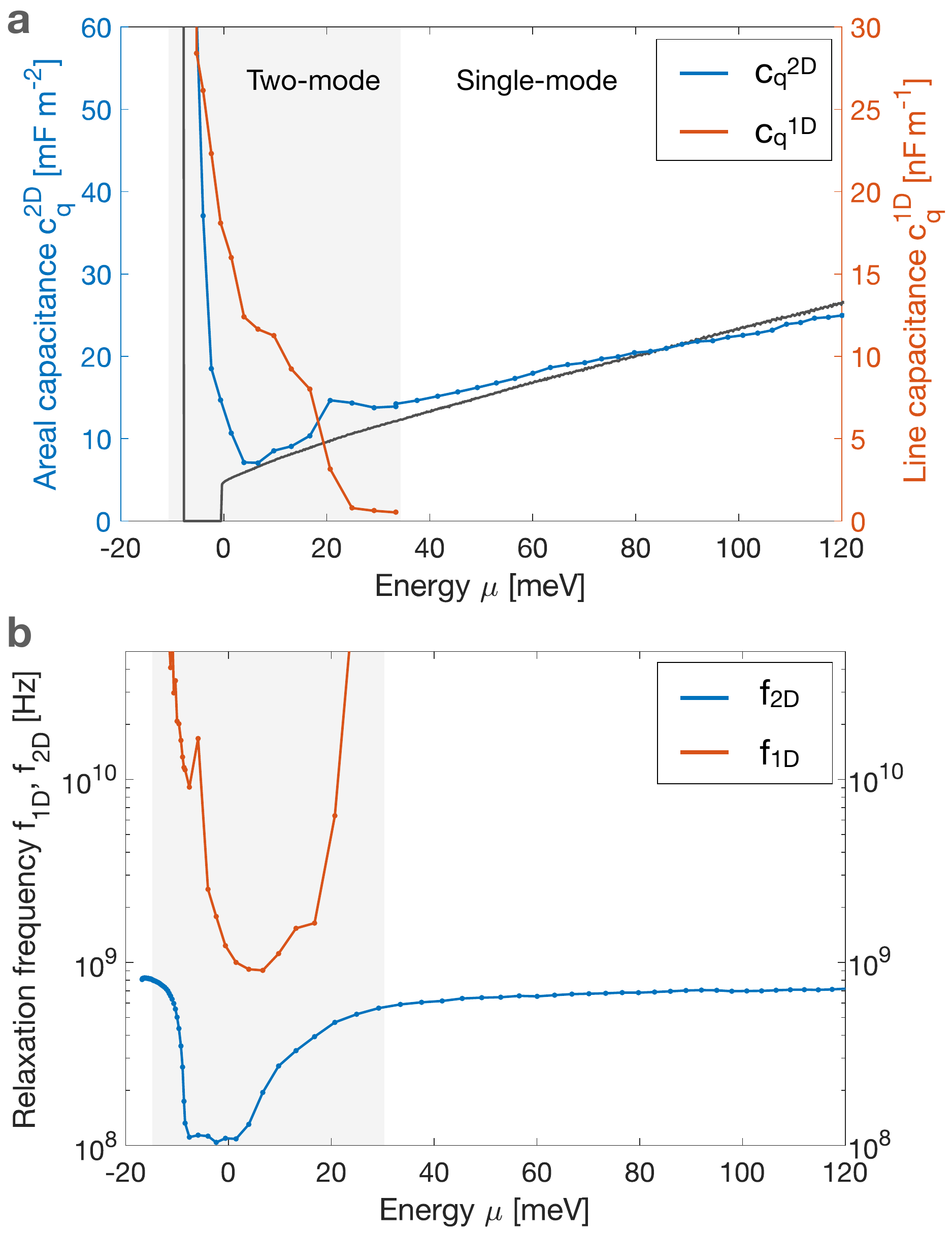}}
\caption{{\bf Quantum capacitances and charge relaxation frequencies:} a) On the left axis, for $L=\SI{5}{\micro\meter}$, the quantum capacitance $c_{\rm q}^{\rm 2D}$ attributed to the 2D bulk carriers is shown as a blue line as a function of energy $\mu$, together with the prediction of the ${\bf k\cdot p}$ calculations (solid grey line). The 1D contribution $c_{\rm q}^{\rm 1D}$ is plotted on the right axis. b) The relaxation frequencies $f_{\rm 2D}$ and $f_{\rm 1D}$) are plotted as function of energy $\mu$, as blue and red line respectively.
}\label{figure:hffitpars}
\end{figure}

The fits obtained with this simple model are in excellent agreement in a wide range of energies corresponding to the valence and conduction bands (as illustrated in Fig.\ref{figure:highfreq}a  and \ref{figure:highfreq}f, see also Supplementary Materials \cite{sm_prl} for linear-scale and Nyquist plots) and confirms the validity of the model. In particular, the crossover between the low-frequency and evanescent regime is highlighted by a clear maximum in Im$\big(Y(\omega)\big)$. The fit yields the value of the HgTe sheet resistance $R$ and of the contact resistance $R_{\rm c}$. $R_{\rm c}$ is found to be relatively independent of $V_{\rm g}$ in the validity range of the model, with $R_{\rm c}=0.3-\SI{4}{\kilo\ohm}$ depending on the device.
However, we observe strong discrepancies arising near the gap between this simple one-carrier model and the data as illustrated in Fig.\ref{figure:highfreq}c, \ref{figure:highfreq}d, \ref{figure:highfreq}e. In particular, the admittance $Y$ exhibits two successive low-frequency regimes (${\rm DC}-\SI{0.1}{\giga\hertz}$ and $0.1-\SI{2}{\giga\hertz}$). We attribute these features to the presence of two independent types of charge carriers, a priori ascribed to the bulk (2D) and edge (1D carriers), with different charge relaxation times. Consequently, we model the system using two parallel distinct distributed RC circuits, each with a contact resistance:
\begin{eqnarray}
Y(\omega)&=&Y^{\rm 1D}(\omega)+Y^{\rm 2D}(\omega)
\end{eqnarray}
where $Y^{\rm 1D/2D}$ are both given by Eq.(\ref{eq:admittance}). We obtain an excellent agreement between model and data over three decades of frequency (Fig.\ref{figure:highfreq}c, \ref{figure:highfreq}d, \ref{figure:highfreq}e) and the two low-frequency regimes are fully captured. The results are shown in Fig.\ref{figure:hffitpars}a. The largest contribution $c_{\rm q}^{\rm 2D}$ is attributed to the two-dimensional bulk states and is observed to follow the ${\bf k\cdot p}$ predictions. The second contribution appears close to the gap, and is consequently attributed to edge carriers $c_{\rm q}^{\rm 1D}$. The above attribution of $c_{\rm q}^{\rm 1D/2D}$ is validated by a separate study of the length dependence of $c_{\rm q}$\cite{sm_prl}, which confirms the values of $c_{\rm q}^{\rm 1D/2D}$, their variations with $\mu$ and more importantly their one- or two-dimensional characters. The one-dimensional component can only be detected in inverted (topological) heterostructures, it is thus likely a signature of the topological edge states. Its order of magnitude $c_{\rm q}^{\rm 1D}\simeq\SI{10}{\nano\farad\per\meter}$ remains however 20 to 50 larger than that of of a bare helical edge state. Though we can not at this stage clearly establish the most plausible explanation, several scenarios can be proposed. First, high quantum capacitances have been measured for quantum Hall edge states \cite{takaoka1994,takaoka1998}. They have been attributed to edge reconstruction under the action of electrostatics and electron-electron interactions \cite{chklovskii1992,chamon1994,wang2017}, leading to the formation of compressible and incompressible stripes. Our observations may also originate in the presence of residual charge puddles near the edges, to which the edge states could couple. These puddles result from disorder, especially when the gap of the QW is small, and may be enhanced on the edges by electrostatics as the topological edge states themselves may screen the action of the gate. Regardless of the microscopic details, a (spin-degenerate) bulk sub-band, could contribute up to \SI{5}{\nano\farad\per\meter}, so that typically 1 to 2 states would reside in the vicinity of the topological edge states, introducing some scattering \cite{vayrynen2014} in the edge states.

Our measurement technique is primarily sensitive to capacitances. Consequently, though the contact and sheet resistances can easily be separated in the single-carrier model, they can only be determined with an accuracy of 50\% in the two-carrier model. We observe nonetheless that, near the gap, the total 2D resistance $R_{\rm t}^{\rm 2D}\gtrsim\SI{40}{\kilo\ohm}$ is much larger than the total 1D resistance $R_{\rm t}^{\rm 1D}\simeq10-\SI{15}{\kilo\ohm}$. In an ideal QSH regime, the geometry resembles that of a mesoscopic capacitor \cite{buttiker1993,gabelli2006,gabelli2012}, and $R_{\rm t}^{\rm 1D}$ should be dominated by a contact resistance, between $\frac{R_K}{4}$ (coherent regime) and $\frac{R_K}{2}$ (incoherent regime)\cite{nigg2008,muller2017}, where $R_K=\frac{h}{e^2}$ is the resistance quantum. Though no quantized resistance plateau is observed in our samples, the extracted values of $R_{\rm t}^{\rm 1D}$ tends to validate the topological origin of the 1D contribution. Higher resistance ($\lesssim R_K$) are sometimes observed, probably due to the presence of scattering as the perimeters of the sample are large ($\simeq\SI{15}{\micro\meter}$ here). This observation justifies a posteriori the description of the edge carriers as a mixed contact and line resistances.

\paragraph{Charge relaxation frequencies --}
This broadband analysis also yields the charge relaxation frequencies given by $f_{\rm 1D/2D}=1/2\pi R_{\rm t}^{\rm 1D/2D}C_{\rm t}^{\rm 1D/2D}$ as plotted in Fig.\ref{figure:hffitpars}b. In the conduction and valence bands, where the single-mode picture holds, we observe that the (bulk) electrons have a high response frequency ($f_{\rm 2D}\simeq0.7-\SI{0.8}{\giga\hertz}$). However, this charge relaxation frequency decreases drastically when the Fermi level approaches the gap and reaches $f_{\rm 2D}\simeq\SI{0.1}{\giga\hertz}$. In the same energy range, the edge electrons appear very mobile, with  $f_{\rm 1D}\simeq\SI{0.9}{\giga\hertz}$. The dynamics thus clearly highlights the presence of mobile edge carriers near the gap in an energy range where bulk carriers are much slower. At high frequency, bulk transport is evanescent over a length $\delta_{\rm 2D}\simeq L\sqrt{\frac{2f_{\rm 2D}}{f}}\ll L$, whereas edge currents penetrate over a larger distance ($\delta_{\rm 2D}\ll\delta_{\rm 1D}\sim L$). The analysis also confirms that topological carriers coexist with the conduction band over a large range of energies ($\simeq\SI{30}{\milli\electronvolt}$), in line with theoretical predictions \cite{dai2008} and observations \cite{nowack2013,bocquillon2016}. 

 The edge and bulk signals are thus clearly distinguished in the microwave spectra, via the charge relaxation frequency, while they are intertwined in measurements of the resistance or quantum capacitance. Beyond this mere identification, the large difference between $f_{\rm 1D}$ and $f_{\rm 2D}$ (about a decade) also has strong implications: it demonstrates that a careful (and geometry-dependent) choice of the drive frequency $f_{\rm 2D}\ll f\ll f_{\rm 1D}$ would allow for addressing dynamically the edge states on time-scales over which the bulk states are frozen. In particular, a favourable situation if found near the bottom of the conduction ($\mu\simeq\SI{0}{\milli\electronvolt}$), where the bulk has a low DOS and is already relatively resistive, and the edge states accessible.

\paragraph{Summary and Outlook --}
As a conclusion, our measurements bring up new information on the transport of topological edge states. While the bulk is mostly insulating in the gap, its DOS is found to remain rather large. Nonetheless, the response of the edge states can be isolated in high-frequency admittance. The edges are rather dense, suggesting that the topological edge states are surrounded by additional states which contribute to the large DOS and to scattering, but not in a significant manner to transport. Their exact nature remains however to be clarified in further studies. Importantly, this work reveals that charge relaxation frequencies for bulk and edge states differ typically by a decade. It implies that microwaves, typically in the GHz range, could be utilized to selectively address and control topological edge transport irrespective of the presence of bulk states. Larger devices with enhanced topological signatures could for example be used to probe for instance Luttinger liquid physics \cite{wu2006,calzona2015,bocquillon2013,kamata2014,muller2017} or dynamical spin and charge transport \cite{inhofer2013,hofer2013,hofer2014}. Finally, we argue that the method is applicable to other topological phases (3D TIs or Weyl semi-metals) where topological transport often compete with numerous less mobile trivial modes.

\section*{Data availability}
The data sets generated and/or analyzed during the current study are available from the corresponding author on reasonable request.

\begin{acknowledgements}
This work has been supported by the ERC, under contract ERC-2017-StG "CASTLES", ERC-2017-Adv "4TOPS", the DFG (SPP 1666, SFB 1170 and Leibniz Program) and the Bavarian Ministry of Education (ENB and ITI). We gratefully acknowledge insightful discussions with D. Bercioux, C. Br\"une, D. Carpentier, M.R. Calvo, T. van der Berg.

\end{acknowledgements}

\section*{Author contributions}

M.C.D., S.H., A.G. and H.Ba. performed the measurements. S.H. and K.B. fabricated the samples, based on MBE layers grown by R.S. and L.L. E.B. supervised the project. All authors participated to the analysis of the results and to the writing of the manuscript.

\newpage

\begin{center}
{{\bf {\large -- Supplementary Information --}}}
\end{center}

\section{Methods}

\subsection{Experimental techniques}

The measurements have been performed in a microwave cryogenic probe station Janis ST-500 equipped with Picoprobe GSG-100 microwave probes. The base temperature is \SI{10}{\kelvin}.
For high-frequency phase-referenced measurements, we use a Vector Network Analyzer (VNA) Anritsu MS4644B. The VNA is first calibrated in the desired frequency range using the SOLT (short-open-load-thru) method on a calibration substrate Picoprobe CS5, thus moving the reference planes to the microwave probe ends. The propagation in the contacts is further corrected by measuring the response of a thru-line. Finally, stray capacitances are subtracted by measuring a dummy device, with identical geometry but without the conductive HgTe mesa structure. The correction from the dummy is found to be slightly too small, as found from control data on non-topological QWs (see Section \ref{section:hf}). A residual stray capacitance of a few fF is often visible. As it is much smaller than $c_{\rm q}^{\rm 1D}$, it does not modify any of our conclusions.

Complementary low-frequency measurements have been made using a lock-in detector (Zurich Instruments HF2LI) combined with a low-noise pre-amplifier. As the setup is not phase-referenced, they only allow for the extraction of the capacitance term (lowest order in $\omega$). Both extract methods agree within the error bars.

\subsection{From $C_{\rm t}$ to $c_{\rm q}$}

\begin{figure*}[ht!]
\centerline{\includegraphics[width=15cm]{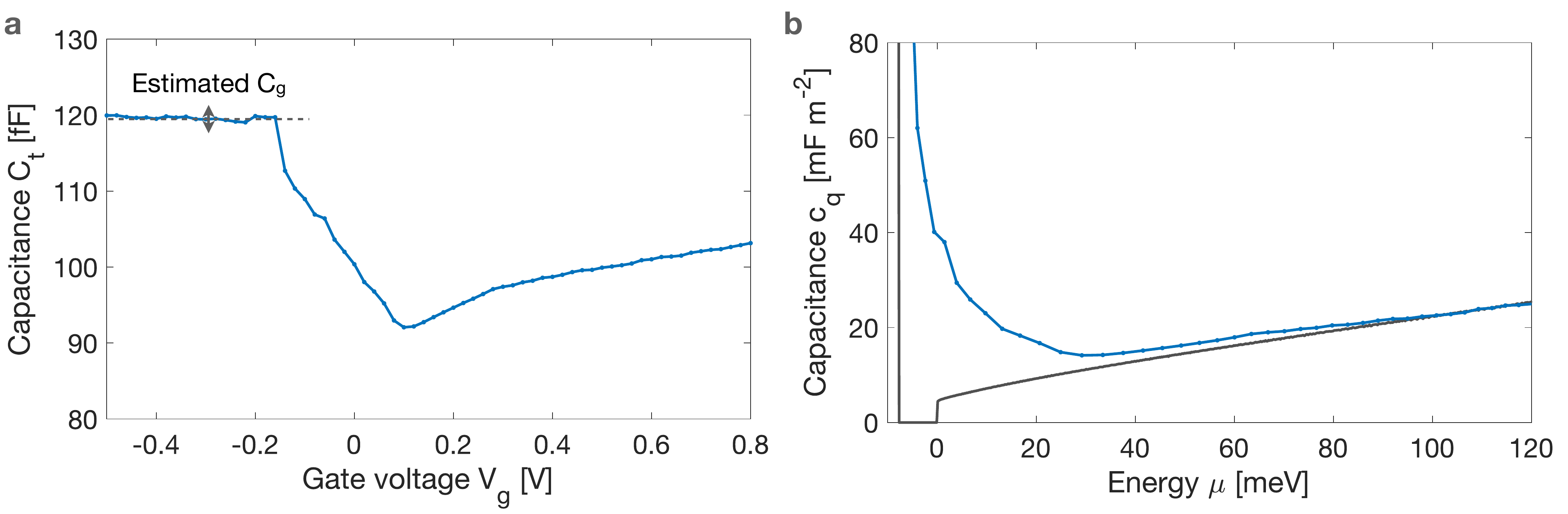}}
\caption{{\bf Evaluation of $C_{\rm g}$:} a) Raw data for $C_{\rm t}$ is plotted in blue, showing a clear minimum attributed to the gap. For negative voltages, $C_{\rm t}$ saturates at $C_{\rm t}\simeq C_{\rm g}$ due to the large DOS in the valence band. This yields $C_{\rm g}$ with a low uncertainty (below 2\%). b) After fine tuning of $C_{\rm g}$ within this 2\% uncertainty margin, a good agreement with ${\bf k\cdot p}$ predictions is found in the conduction band regime.
}\label{figure:SI_cg}
\end{figure*}

The measurements with a lock-in amplifier or a VNA directly yields the total capacitance $C_{\rm t}$ of the sample. An example measurement of $C_{\rm t}$ as function of $V_{\rm g}$ is shown in Fig.\ref{figure:SI_cg}a. The gap feature is immediately visible around $V_{\rm g} \simeq\SI{0.1}{\volt}$ as a dip in $C_{\rm t}$. The contrast is very good (typically 25 \% modulation of $C_{\rm t}$) thanks to the very thin insulator layer that maximizes the value of $C_{\rm g}$. A clear saturation is observed for $V_{\rm g}< \SI{-0.2}{\volt}$ that signals the very heavy valence band, for which $C_{\rm t}\simeq C_{\rm g}$. From the saturation value, we obtain $C_{\rm g}$ with an estimated error $<2\%$. Knowing $C_{\rm g}$, we compute $C_{\rm q}$ and $c_{\rm q}=\frac{C_{\rm q}}{S}$, where $S$ is the area of the sample. To avoid any systematic error in $c_{\rm q}$, we measure $S$ with microscope pictures. In the length dependence (see next section), we use $\sqrt{S}$ rather than the nominal value $L$.
Despite the accuracy of less than $2\%$ on $C_{\rm g}$, the agreement with ${\bf k\cdot p}$ prediction can sometimes be insufficient. We take these theoretical predictions as a reference and thus fine tune the value of $C_{\rm g}$ (within $2\%$) to obtain the best agreement, and finally obtain the plot of  $c_{\rm q}$ as function of $\mu$ as presented in Fig.\ref{figure:SI_cg}

\subsection{Single mode high frequency admittance}

Following earlier works \cite{pallecchi2011,inhofer2017}, we model the capacitor as a distributed RC line (of length $L$ along the $x$ axis, see Main text). On the source side, an oscillating voltage $V_{\rm ac}(t)$ is applied, while a dc gate voltage is set to the other end of the line. Solving the Kirchhoff equations, one can easily show that the local voltage $V(x)$ and current $I(x)$ read :
\begin{eqnarray}
V(x)&=&V_+{\rm e}^{i\gamma x}+V_-{\rm e}^{-i\gamma x}\\
I(x)&=&\frac{i\gamma}{r}\left(V_+{\rm e}^{i\gamma x}-V_-{\rm e}^{-i\gamma x}\right)
\end{eqnarray}
with $\gamma=\sqrt{i\omega \mathcal{R} \mathcal{C}_{\rm t}}$. Solving for the above mentioned boundary conditions, this immediately yields the admittance $Y_0(\omega)$ of the line:
\begin{eqnarray}
Y_0(\omega)&=&\frac{I(0)}{V(0)}\,=\,\frac{i\gamma}{r}\tanh(i\gamma L)\,=\,\frac{ik}{R}\tanh\left(ik\right)
\end{eqnarray}
with $k=\sqrt{i\omega R C_{\rm t}}$, $R= \mathcal{R}L, C_{\rm t}= \mathcal{C}_{\rm t}L$.
Taking into account the contact resistance between the QW and the gold lead of the source, a discrete resistance $R_{\rm c}$ is further attached to one end and the admittance $Y$ is finally given by:
\begin{eqnarray}
Y(\omega)&=&\frac{1}{R_{\rm c}+ 1/Y_0(\omega)}\,=\,\frac{1}{R_{\rm c}+ \frac{R}{ik}\frac{1}{\tanh\left(ik\right)}}
\end{eqnarray}
The low frequency expansion of $Y(\omega)$ can then be calculated, and gives:
\begin{eqnarray}
Y(\omega)&=&i\omega C_{\rm t}+\omega^2R_tC_{\rm t}^2+o(\omega^2)
\end{eqnarray}
with $R_{\rm t}=R_{\rm c}+\frac{R}{3}$.

\section{Low frequency and length study}
\label{section:lf}

The following experiment in the low-frequency range has been repeated on approximately 30 devices, fabricated from 3 epilayers. Systematic behavior has been observed, which supports the high-frequency analysis reported in the main text. We here present their conclusions.  The sample presented in the main text is here denoted as Topological B.

\subsection{Low-frequency $RC$-like behavior}

This paragraph focuses on the low-frequency quantities $R_{\rm t}$, $C_{\rm t}$ and $c_{\rm q}$ (per unit area) measured on a non-topological sample and two topological ones, including the one presented in the main text. In particular, it establishes, through the comparison of capacitors of different sizes, the existence of a 1D density of states in topological samples, as expected due to the presence of topological edge states. The capacitance values extracted for the edge states are in excellent agreement with the ones coming from the high frequency measurements and validate the high frequency analysis.

\begin{figure}[ht]
\centerline{\includegraphics[width=8.5cm]{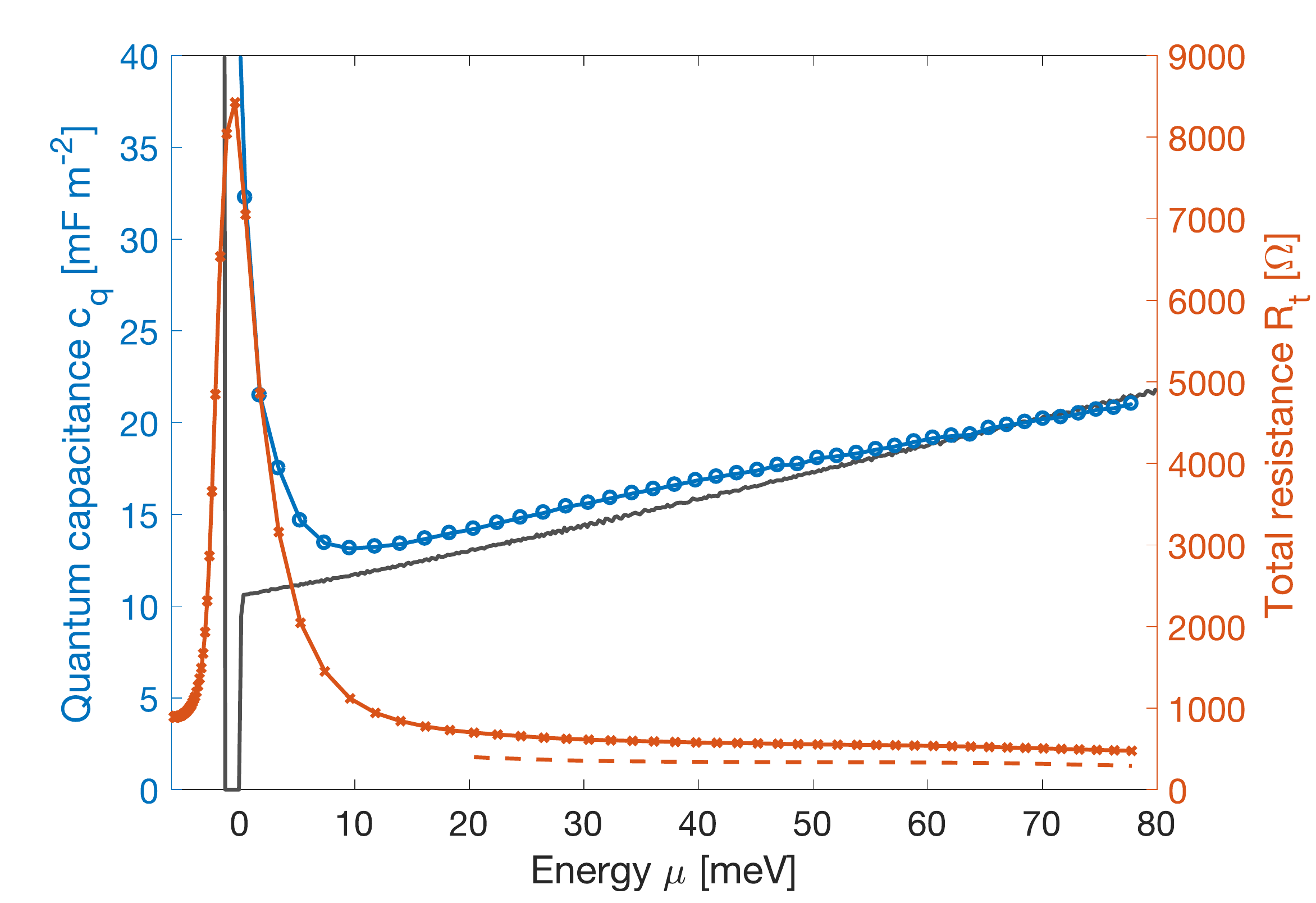}}
\caption{{\bf Quantum capacitance and resistance of a topological HgTe quantum well:} The quantum capacitance per unit area $c_{\rm q}$ (blue line, left axis) and the total resistance $R_{\rm t}$ (red line, right axis) are plotted as a function of energy $\mu$. As a grey line, the computed value of $c_{\rm q}$ given the ${\bf k\cdot p}$ band structure is also reproduced. As a dashed red line, the contact resistance $R_{\rm c}$ is plotted in the energy range in which it can meaningfully be extracted from the high-frequency analysis. This dataset has been measured on Topological A for $L=\SI{10}{\micro\meter}$.
}\label{figure:CqR}
\end{figure}

In Fig.\ref{figure:CqR}, the total resistance $R_{\rm t}$ of Topological A is presented together with the quantum capacitance $c_{\rm q}$. In the valence and conduction band, $R_{\rm t}$ is very low as the QW is very conductive, and saturates close to the value of the contact resistance, $R_{\rm t}\simeq R_{\rm c}$ (the latter being extracted from the high-frequency analysis, and depicted as the dashed line). The gap is clearly indicated by a sharp peak in $R_{\rm t}$. For trivial samples (not shown), the gap is quite insulating and $R_{\rm t}>\SI{400}{\kilo\ohm}$. As stated in the main text, for topological samples in an ideal QSH regime, the geometry resembles that of a mesoscopic capacitor \cite{buttiker1993,gabelli2006,gabelli2012}, and a charge relaxation resistance between $\frac{R_K}{4}$ (coherent regime) and $\frac{R_K}{2}$ (incoherent regime) is expected\cite{nigg2008,muller2017}, where $R_K=\frac{h}{e^2}$ is the resistance quantum. Though many measured peak resistances lie in this range (see Fig.\ref{figure:CqR}), none of the samples however exhibit a clear quantized resistance plateau, and higher values ($\sim R_K$) are sometimes reached, probably due to the presence of scattering as the perimeters of the sample are large (\SI{30}{\micro\meter} here).
Importantly, we found that in all samples the minimum of $c_{\rm q}$ and the maximum of $R_{\rm t}$ are not aligned. This indicates that the gap hosts a large amount of bulk states that do not participate very efficiently to transport, and that the Fermi energy is mostly pinned to the valence band. In contrast, the minimum of $c_{\rm q}$ (here around $\mu\simeq\SI{10}{\milli\electronvolt}$) corresponds to a smaller DOS, but a higher conductance.

\begin{figure*}[ht!]
\centerline{\includegraphics[width=\textwidth]{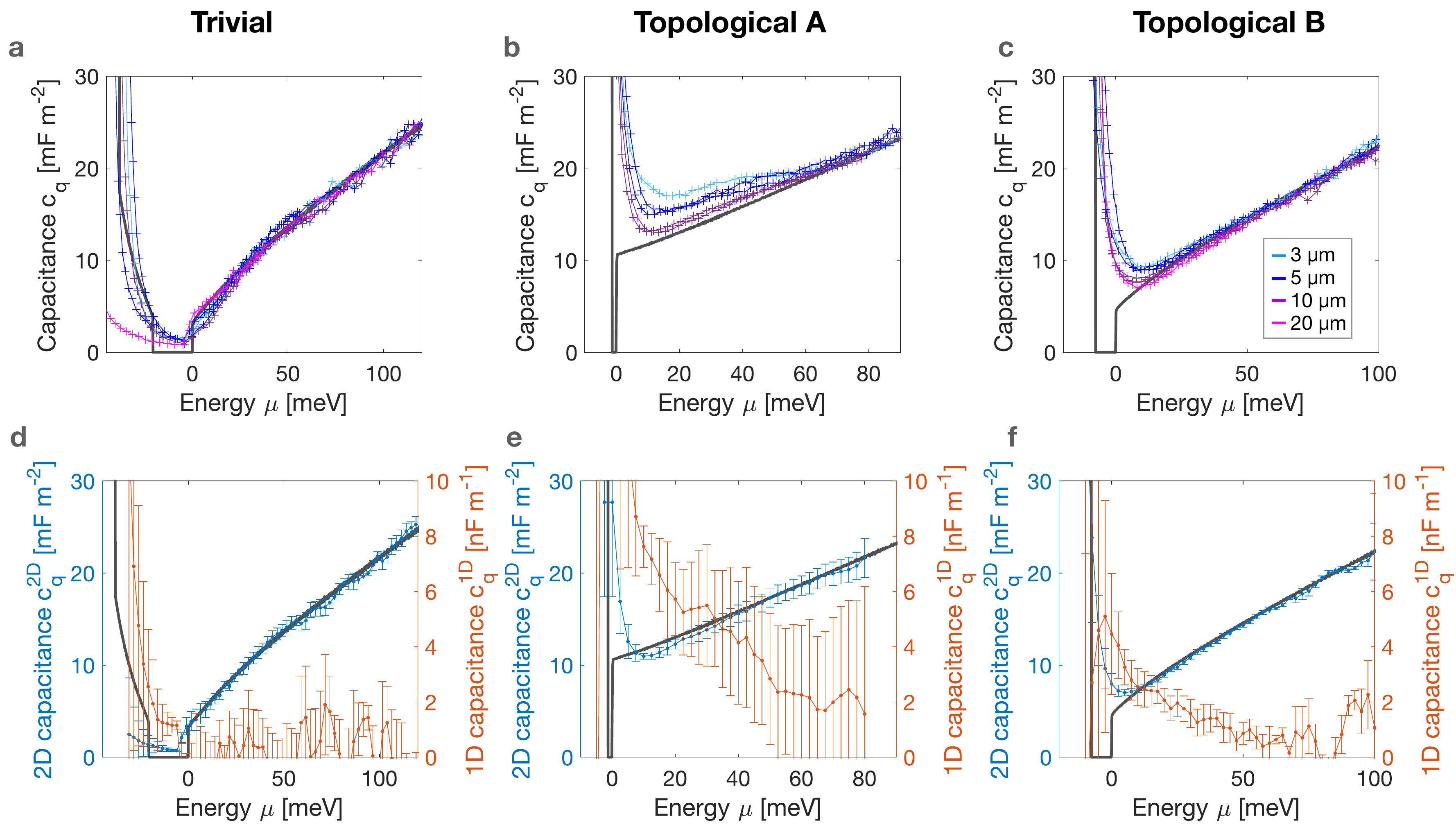}}
\caption{{\bf Length dependence of the quantum capacitance:} a, b, c) Quantum capacitance $c_{\rm q}$ (per unit area) as function of energy $\mu$ measured for different lengths $L=3,5,10, \SI{20}{\micro\meter}$. The grey line corresponds to the computed value of $c_{\rm q}$ given the ${\bf k\cdot p}$ band structure, the colored lines correspond to experimental data for a size $L$ indicated by the caption d, e, f) On the left axis, two-dimensional component $c_{\rm q}^{\rm 2D}$ (per unit area) as a blue line, the grey line corresponds to $c_{\rm q}$ computed from the ${\bf k\cdot p}$ band structure. On the right axis, one-dimensional component $c_{\rm q}^{\rm 1D}$ (per unit length) as a red line.
}\label{figure:length}
\end{figure*}

\subsection{Length dependence}
\label{section:lf:length}

As illustrated in Fig.\ref{figure:length}a, \ref{figure:length}b, \ref{figure:length}c, for all samples the curves are superimposed for $\mu>\SI{60}{\milli\electronvolt}$ signaling that $c_{\rm q}$ does not depend on $L$, and the system is thus two-dimensional. However, a clear length dependence appears for lower energies in the topological samples (Fig.\ref{figure:length}b and \ref{figure:length}c). Assuming the existence of edge channels of a finite width, we can decompose the quantum capacitance contribution in two parts:
\begin{itemize}
    \item a lineic contribution coming from the edge modes scaling as $S_1=3WL-2W^2\simeq 3WL$, where $W$ is the effective width of the edge modes.
    \item a surfacic contribution coming from the bulk modes scaling as $S_2=(L-W)(L-2W)=S_t-S_1$
\end{itemize}

The edge states are assumed to run on three sides only of the gated surface $L^2$, but not on the side nearby the contact.
We assume that the geometric capacitance $C_{\rm g}=c_{\rm g}L^2$ is in series with the two quantum capacitances $c_{\rm q1} S_1$ and $c_{\rm q2} S_2$ in parallel:
\begin{eqnarray}
C_{q}&=&c_{\rm q1} S_1+c_{\rm q2} S_2\\
&=&c_{\rm q}^{\rm 2D}L^2+3 c_{\rm q}^{\rm 1D}L
\end{eqnarray}
with $c_{\rm q}^{\rm 2D}=c_{\rm q1}$ and $c_{\rm q}^{\rm 1D}=W(c_{\rm  q2}-c_{\rm q1})$.

Fits of the data for each $\mu$ finally yields $c_{\rm q}^{\rm 1D}$ and $c_{\rm q}^{\rm 2D}$, plotted in Fig.\ref{figure:length}e and \ref{figure:length}f. The two-dimensional part $c_{\rm q}^{\rm 2D}$ follows the ${\bf k\cdot p}$ calculations, confirming the validity of the modeling. In contrast, the one-dimensional contribution $c_{\rm q}^{\rm 1D}$ appears at energies $\mu<\SI{40}{\milli\electronvolt}$ and is at its largest near the gap. The order of magnitude $c_{\rm q}^{\rm 1D}\simeq 5-\SI{10}{\nano\farad\per\meter}$ is however 25 to 50 times larger than that expected for a merely one-dimensional edge (typ. \SI{0.2}{\nano\farad\per\meter}), in agreement with the values extracted from the two-fluid model analysis presented in the main text. In contrast, the Trivial sample does not show any clear linear one-dimensional contribution, confirming the absence of edge transport in the gap (Fig.\ref{figure:length}a and \ref{figure:length}d). We however note the scaling analysis can hardly be performed for negative energies. Indeed, $c_{\rm q}$ varies abruptly, and errors and variations in the determination of $C_{\rm g}$  are amplified in this regime especially for Trivial as the QW is strongly depleted in the gap.

The above modelling can be used to link the 1D and 2D components of the total capacitance to the two components of the quantum capacitance, using the assumption $\frac{3 L c_{\rm q}^{\rm 1D}}{c_{\rm q}^{\rm 2D}L^2}\ll 1$,
\begin{eqnarray}
C_{\rm t}&=&\frac{C_{\rm q}C_{\rm g}}{C_{\rm g}+C_{\rm q}}\\
&=&c_{\rm t}^{\rm 2D}L^2+3 c_{\rm t}^{\rm 1D}L
\end{eqnarray}
with $c_{\rm t}^{\rm 2D}=\frac{c_{\rm q}^{\rm 2D}c_{\rm g}}{c_{\rm g}+c_{\rm q}^{\rm 2D}}$ and $ c_{\rm t}^{\rm 1D}= c_{\rm q}^{\rm 1D}\left(\frac{c_{\rm t}^{\rm 2D}}{c_{\rm q}^{\rm 2D}}\right)^2$.

\section{Additional microwave data}
\label{section:hf}

\subsection{Linear-scale and Nyquist representation of microwave spectrums}

Fig.\ref{figure:highfreq} displays two spectrums from Fig.4 (main text), respectively taken at $\mu=\SI{-17}{\milli\electronvolt}$ and $\mu=\SI{-2.4}{\milli\electronvolt}$. They are first shown using a linear frequency scale (panels a and b) to emphasize the low-frequency linear/parabolic behavior, expected for a RC circuit, as well as and the good quality of the fits at high frequency. Panels c and d show the same data in a Nyquist representation. The single-mode behavior (c) appears as a half-circle typical of the RC circuit response, with deviations appearing at higher frequency when entering the evanescent regime. The two-mode response (d) shows the succession of two arcs.


\begin{figure*}[ht!]
\centerline{\includegraphics[width=13cm]{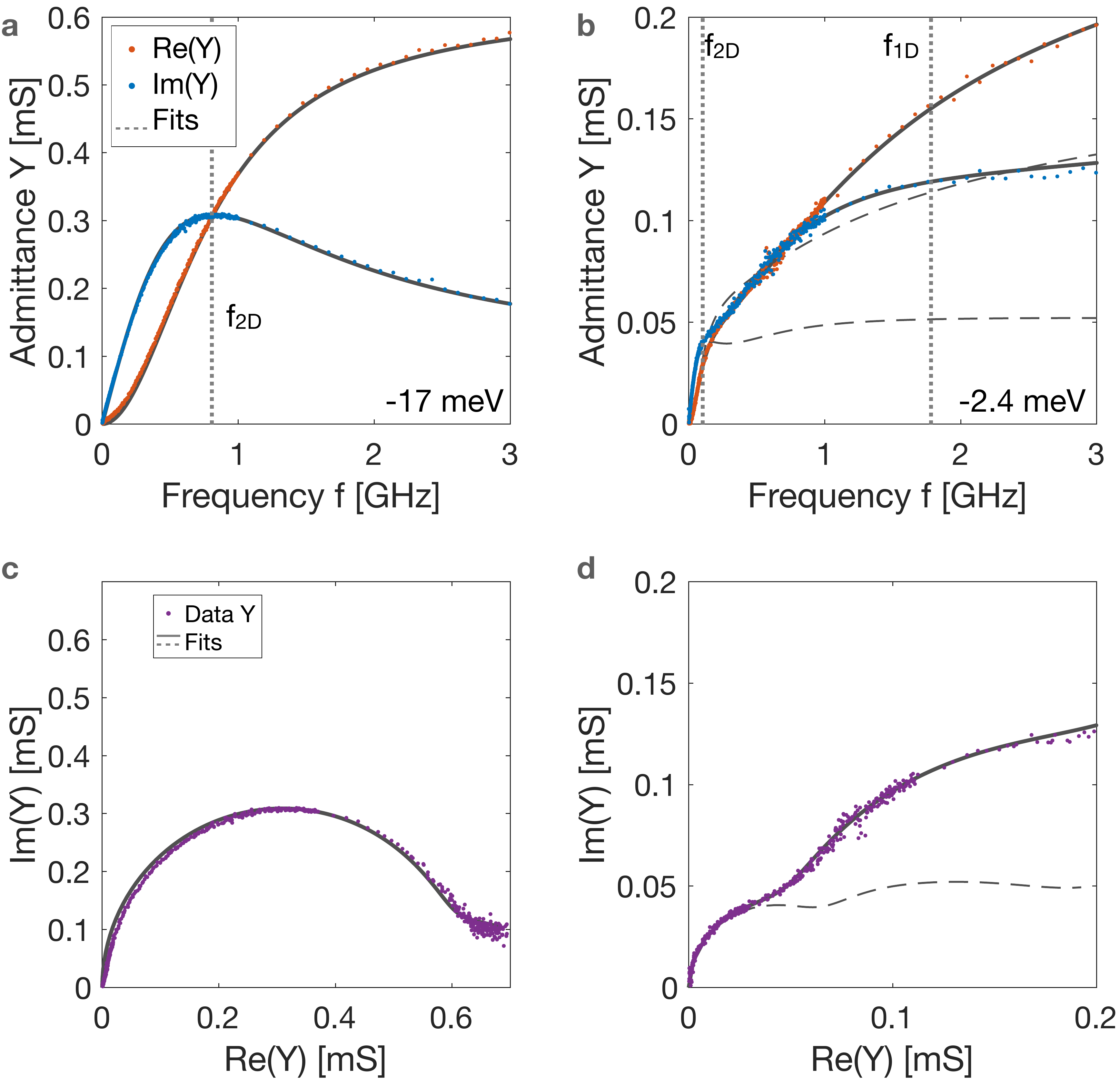}}
\caption{{\bf Microwave admittance $Y$:} a and b) Admittance $Y$ as function of frequency, measured at $\mu=\SI{-17}{\milli\electronvolt}$ (resp. and $\mu=\SI{-2.4}{\milli\electronvolt}$) for $L=\SI{5}{\micro\meter}$. c and d) Same data plotted as a Nyquist representation in the complex plane. The best fits are plotted as dark grey lines. For panels c and d, the single-mode results appear as dashed lines.}
\label{figure:highfreq}
\end{figure*}

\subsection{Control data on a non-topological quantum well}

\begin{figure*}[ht!]
\centerline{\includegraphics[width=19cm]{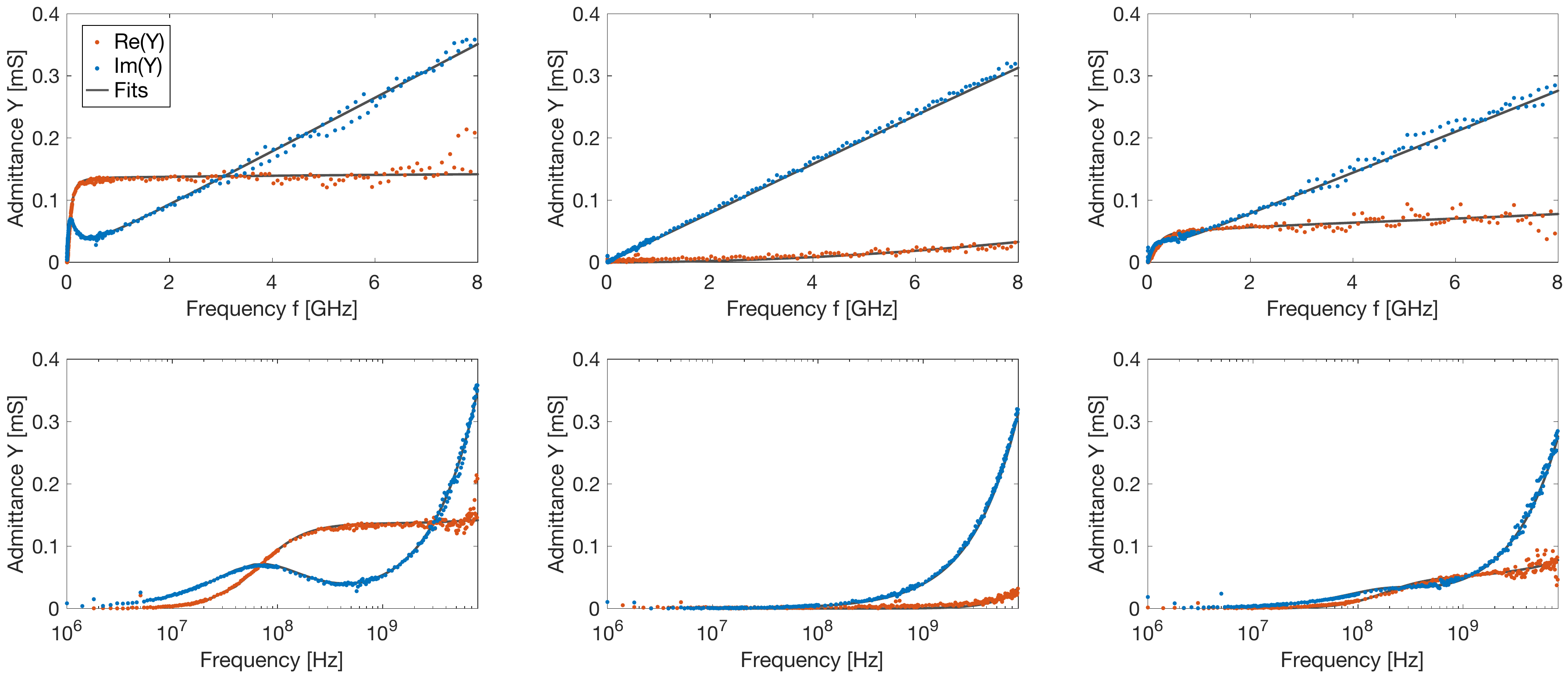}}
\caption{{\bf Microwave admittance $Y$ of a non-topological device:} a,b,c) Admittance $Y$ as function of frequency on a linear scale, taken in the gap region of the Trivial sample, for $L=\SI{10}{\micro\meter}$. d,e,f) Same data plotted using a logarithmic frequency scale.
}\label{figure:trivial_rf}
\end{figure*}

High frequency data obtained using the reference sample Trivial (see Section \ref{section:lf}) are shown on Fig.\ref{figure:trivial_rf}. Outside the gap region (not shown), the samples are nicely fitted by a single-mode response. Near the gap, the resistance is very large as the sample become quite insulating, with resistance reaching several hundreds of \si{\kilo\ohm}. A second feature becomes visible near the resistance peaks, as the admittance becomes very small. It is characterized by a linear increase of Im($Y$), while Re($Y$) remains almost constant.

This linear behavior of Im($Y$) is characteristic of an additional capacitive coupling, which we model again with our two-mode model. For the following reasons, we attribute it to a residual stray capacitance, not perfectly corrected by the dummy structure:
\begin{itemize}
\item {\sl Order of magnitude}: the residual capacitance is on the order of a few fF, representing typically 20 to 50 \% of the dummy correction. It is in any case more than 10 times smaller than the observed 1D mode in topological samples. 
\item {\sl Associated resistance}: the resistance associated to this mode is very low ($<\SI{1}{\ohm}$), and is incompatible with any transport channel in a semi-conductor device.
\item {\sl Gate dependence}: as visible on the different panels of Fig.\ref{figure:trivial_rf}, the slope (hence the capacitance) is almost gate independent.
\item {\sl Calibration dependence}: data sets with different calibration show different values of the residual capacitance, substantiating the possibility of an inaccurate dummy correction.
\end{itemize}

Remarkably, on panel b, the bulk response has totally vanished, and the stray capacitance thus dominates. This study actually gives a good estimate of the measurement accuracy, typically on the order of a few fF. It confirms the validity of the length and high-frequency studies, as the variations attributed to $c_{\rm q}^{\rm 1D}$ are typically one or two orders of magnitude larger.

\subsection{Study of a larger topological device ($L=\SI{10}{\micro\meter}$)}

\begin{figure*}[ht!]
\centerline{\includegraphics[width=15cm]{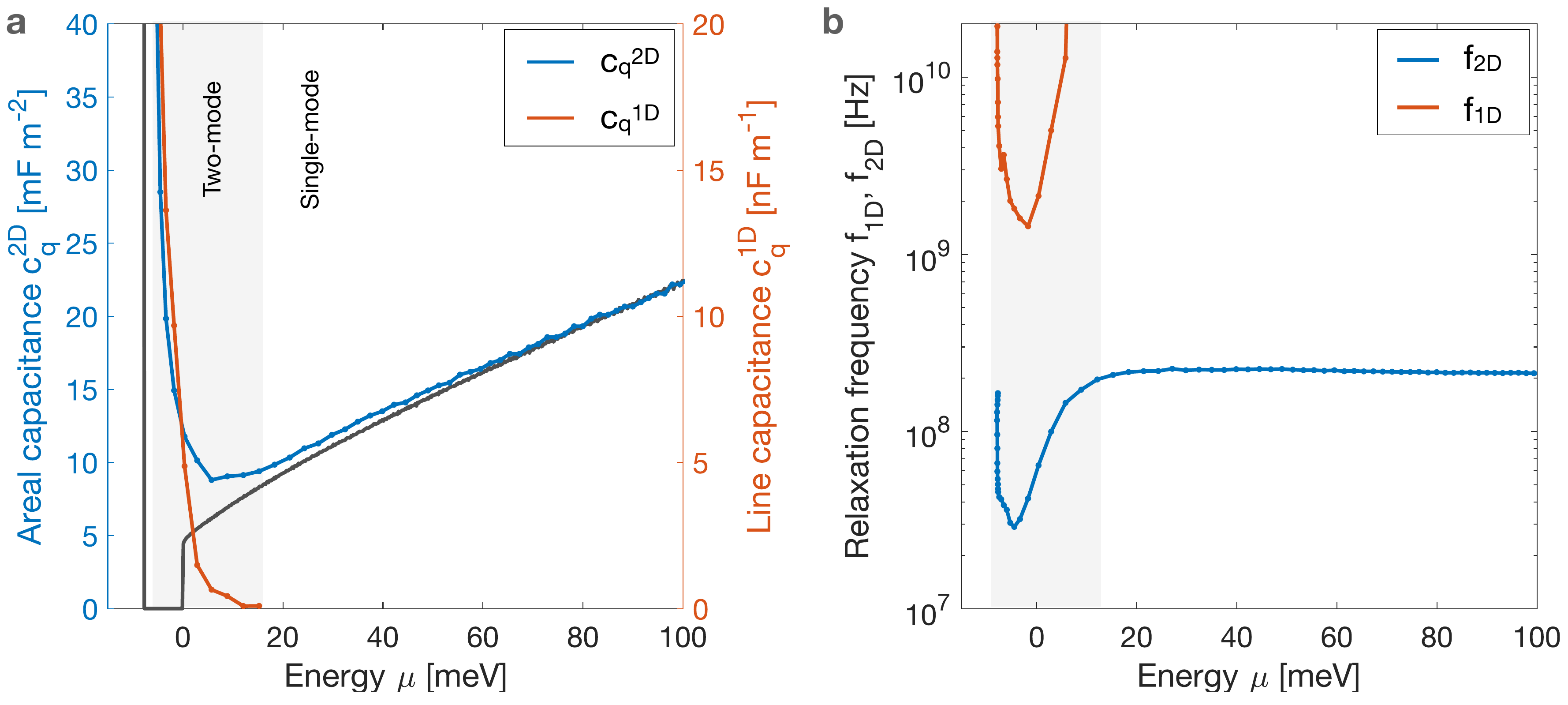}}
\caption{{\bf Quantum capacitances and charge relaxation frequencies for $L=\SI{10}{\micro\meter}$} a) On the left axis, the quantum capacitance $c_{\rm q}^{\rm 2D}$ attributed to 2D bulk carriers is shown as a blue line as a function of energy $\mu$, together with the prediction of the ${\bf k\cdot p}$ calculations (solid grey line). The 1D contribution $c_{\rm q}^{\rm 1D}$ is plotted on the right axis. b) The relaxation frequencies $f_{\rm 2D}$ and $f_{\rm 1D}$) are plotted as function of energy $\mu$, as blue and red line respectively. 
}\label{figure:highfreq_c10}
\end{figure*}

In this section, we show additional data obtained on a larger device than the one presented in the main text, namely $L=\SI{10}{\micro\meter}$. The single- and two-mode fits are performed similarly and the results are presented in Fig.\ref{figure:highfreq_c10}, yielding consistent data with the device shown in the main text ($L=\SI{5}{\micro\meter}$).

The trends previously observed for $c_{\rm q}^{\rm 2D}$ and $c_{\rm q}^{\rm 1D}$ are globally confirmed, as well as the corresponding magnitudes, thus validating the study shown in the main text. Interestingly, one observes that the 2D charge relaxation frequency saturates at $f_{\rm 2D}\simeq \SI{200}{\mega\hertz}$. From this data, one can calculate the diffusion constant using the Einstein relation \cite{pallecchi2011}, which remains remarkably constant over the whole conduction band:
\begin{equation}
    D_{2D}=\frac{1}{R_{\rm 2D}c_{\rm q}^{\rm 2D}}\simeq \SI{0.03}{\meter\square\per\second}
\end{equation}
As $R_{\rm 2D}$ is here the sheet resistance per square and $c_{\rm q}^{\rm 2D}$ is the capacitance is per unit area, $D_{2D}$ should be independent of $L$. Calculating $D_{2D}$ for the sample presented in the main text, we indeed obtain the same value.

\bibliography{BibCapa2D.bib}

\end{document}